# The green hydrogen ambition and implementation gap

Adrian Odenweller[1,2], Falko Ueckerdt[1]

[1] Potsdam Institute for Climate Impact Research, Member of the Leibniz Association, Potsdam, Germany

[2] Global Energy Systems Analysis, Technical University of Berlin, Berlin, Germany

## Abstract

Green hydrogen is critical for decarbonising hard-to-electrify sectors, but faces high costs and investment risks. Here we define and quantify the green hydrogen ambition and implementation gap, showing that meeting hydrogen expectations will remain challenging despite surging announcements of projects and subsidies. Tracking 137 projects over three years, we identify a wide 2022 implementation gap with only 2% of global capacity announcements finished on schedule. In contrast, the 2030 ambition gap towards 1.5°C scenarios is gradually closing as the announced project pipeline has nearly tripled to 441 GW within three years. However, we estimate that, without carbon pricing, realising all these projects would require global subsidies of $1.6 trillion ($1.2 – 2.6 trillion range), far exceeding announced subsidies. Given past and future implementation gaps, policymakers must prepare for prolonged green hydrogen scarcity. Policy support needs to secure hydrogen investments, but should focus on applications where hydrogen is indispensable.

## Introduction

There is a widespread consensus among scientists[1–5], industry[6] and increasingly also policymakers[7] that green hydrogen, produced from renewable electricity via electrolysis, is critical for reducing emissions in end-use applications that defy straightforward electrification. Additionally, hydrogen is a promising candidate for long-duration energy storage of renewables[8,9] and the precursor to all electrofuels[10], which are highly versatile yet costly[11]. Consequently, policy measures to stimulate the hydrogen market ramp-up are gaining momentum as more than 40 governments have already adopted hydrogen strategies[1,7]. Prominent examples are the supply-side subsidies implemented through the US Inflation Reduction Act[12] and the EU Hydrogen Bank[13]. Such policy support is urgently required: to meet the median ambition in 1.5°C scenarios, 350 GW by 2030, green hydrogen production needs to grow at least 190-fold, more than doubling each year. However, implementation is not going according to plan.

Following a surge of enthusiasm[14,15], the green hydrogen market and associated expectations have recently entered a phase of consolidation[16] as high costs[17], limited demand[18] and lagging implementation of support policies[1] are hampering deployment. Shortfalls in the announced deployment of electrolysers, the key component for green hydrogen production, are representative of the systemic challenges of scaling up supply, demand and infrastructure at the same time. In 2022, instead of the 3.8 GW electrolysis capacity initially expected, eventually only 0.63 GW was realised on time (Figure 1a). Similarly, in 2023, of the 9.4 GW expected, only 1.86 GW will be realised at best. In stark contrast to these recent setbacks, announced future growth rates of green hydrogen have increased substantially over the past three years, indicating a backlog of projects as well as further increasing ambition (Figure 1b). This raises questions such as (i) can recent failure rates and the looming "valley of death"[19] be overcome to meet updated project announcements, (ii) did the expected role of hydrogen in ambitious climate change mitigation scenarios change, and (iii) what are plausible implementation pathways under currently announced hydrogen support policies?

In this paper, we structure and analyse the past and future challenges of the nascent green hydrogen industry by introducing and quantifying the green hydrogen ambition and implementation gap. This builds on the well-established concepts of emission gaps[20] and recent extensions towards a carbon dioxide removal gap[21]. Looking back, we define the past implementation gap as the difference between announced and eventually realised project capacity in 2022 and 2023 (Figure 1a, ①). Looking ahead to 2030, we define the ambition gap as the difference between 1.5°C scenario requirements and announced projects, and find that it has been gradually closing in the past three years (Figure 1b, ②). However, on the other hand, this has been accompanied by a widening future implementation gap, which we define as the difference between announced projects and projects that are backed by policies in 2030 (Figure 1b, ③). Analysing the competition of green hydrogen with natural gas, we estimate that green hydrogen will require subsidies, or alternative policies such as end-use quotas, for at least another decade – even with ambitious carbon pricing and much longer without. This article is structured along these three gaps, followed by a discussion of policy implications in order to safeguard climate targets against uncertain green hydrogen supply.

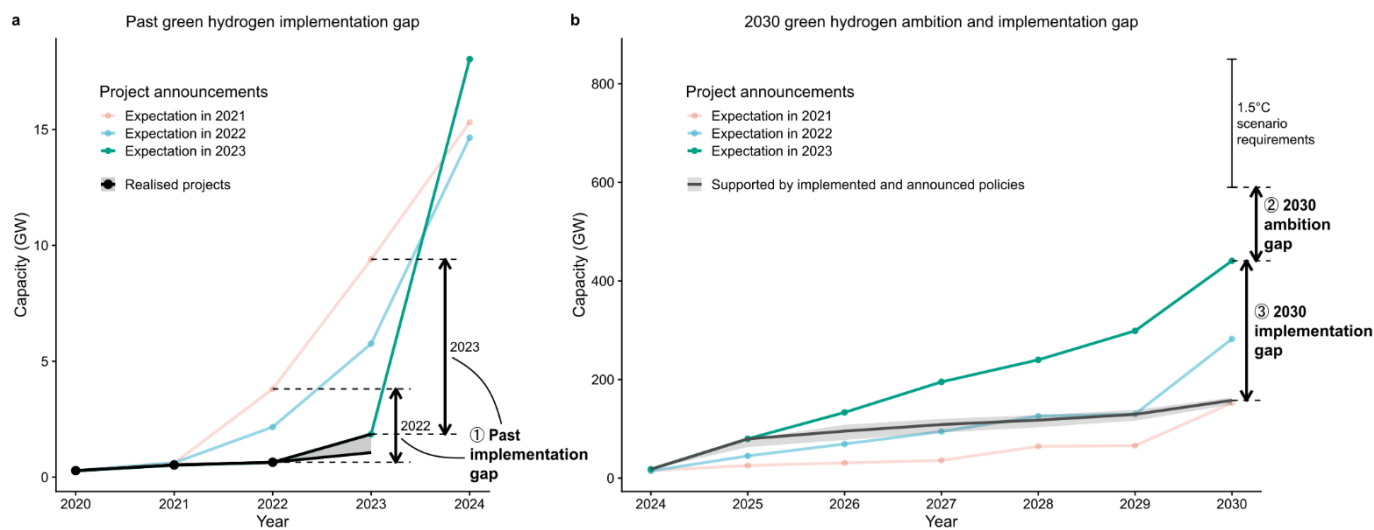

*Figure 1: The green hydrogen ambition and implementation gap in the past and the future.* **a,** Past green hydrogen implementation gap in 2022 and 2023, defined as the difference between project announcements and realised projects (see Figure 3). **b,** Green hydrogen ambition and implementation gap in 2030. We define the 2030 ambition gap as the difference between 1.5°C scenarios and project announcements (see Figure 4b). The depicted scenario range shows the IEA Net Zero Emissions by 2050 Scenarios, while the full analysis includes further scenarios (see Figure 4a, Extended Data Figure 2 and Methods). We define the 2030 green hydrogen implementation gap as the difference between project announcements and our estimate of projects that are supported by implemented demand-side policies or can be financially supported with currently announced subsidies (see Figure 5 and Extended Data Figure 9). The grey line indicates our central estimate, while the light grey corridor indicates the uncertainty range spanned by the sensitivity analysis (see Methods). Green hydrogen project announcements are shown in terms of electrical input capacity of the electrolyser.

## The wide green hydrogen implementation gap in 2022 and 2023

Green hydrogen project announcements reveal two opposing trends over the past three years. First, there has been a notable short-term setback, with expectations of realised capacity diminishing as projects approach their announced launch year (Figure 2a). This trend of downward-adjusted expectations persists in both 2022 and 2023, indicating a dramatic green hydrogen implementation gap in recent years. Second however, this trend reverses from 2024 onwards, with project announcements increasing steadily over the past three years (Figure 2b). The announced steep mid-term growth is mostly driven by the EU, which accounts for the largest share of announced capacity by 2030, followed by Australia, and Central and South America (Figure 2d). These opposing trends raise the question whether future promises can overcome past setbacks. We address this question in the next section, following the quantification of the 2022 and 2023 green hydrogen implementation gap.

Tracking 137 individual green hydrogen projects announced globally for 2022 over the past three years (see Methods), we observe a substantial implementation gap (Figure 3a). A comparison of expectations in 2021 with the final outcome reveals that only 2% of the total capacity announced for 2022 was realised on time, with 42% experiencing delays and 56% disappearing altogether (Figure 3b). Even within the subset of projects that had secured final investment decisions (FIDs) or were already under construction, only 15% were completed on time. Similarly, comparing expectations in 2022 with the final outcome, merely 4% of total announced capacity was realised on time, with 71% facing delays and 25% disappearing (Figure 3c). Notably, projects in the feasibility study or concept stage had a success rate of zero, implying that projects lacking an FID at the time of announcement were never realised on time. These high failure rates are not compensated by the influx of newly announced projects or projects that were delayed from previous years (grey bars in Figure 3a) such that a dramatic green hydrogen implementation gap of more than 3 GW, equivalent to more than 95% of initially announced capacity, remains in 2022.

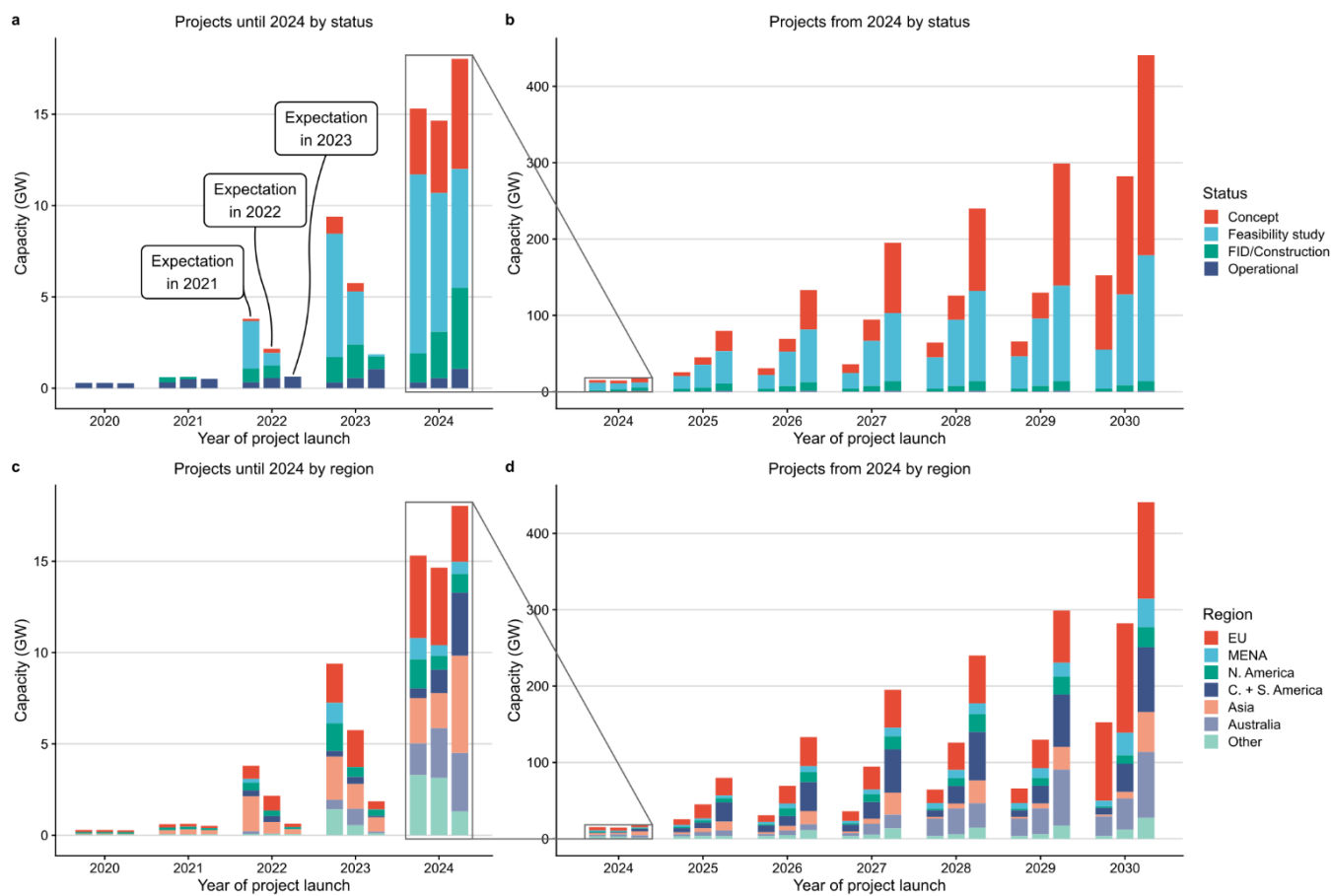

*Figure 2: Green hydrogen project announcements as expected in 2021, 2022 and 2023. **a-b**, Project announcements by status from 2020-2024 and 2024-2030, respectively. FID stands for final investment decision. **c-d**, Project announcements by region from 2020-2024 and 2024-2030, respectively. For each year of project launch there are three bars. The left bar shows the expectation in 2021, the middle bar shows the expectation in 2022, and the right bar shows the expectation in 2023, each of which correspond to different project database snapshots (see Methods). Two main trends are visible. First, in 2022 and 2023 project announcements decrease strongly as the year of project launch approaches (a, c), leading to a wide green hydrogen implementation gap (see Figure 3). Second, after 2024, this pattern reverses as the project pipeline has surged over the past three years (b, d), thereby gradually closing the green hydrogen ambition gap to 1.5°C scenarios (see Figure 4). However, the vast majority of projects have not secured an FID yet (b), which gives rise to the 2030 green hydrogen implementation gap due to a mismatch of required and announced policies (see Figure 5).*

The high failure rates of green hydrogen projects in 2022 could be linked to supply chain disruptions caused by the COVID-19 pandemic and surging electricity prices during the European energy crisis. However, tracking 207 individual projects announced for 2023 reveals similar trends (Extended Data Figure 1). For 2023, some uncertainty remains since the database is released in October, meaning the final outcome of project announcements can only be determined in the following year (see Methods). Despite this uncertainty, already by October 2023 expectations had dropped substantially, indicating that no more than 14% of the total capacity announced for 2023 would be realised on time (Extended Data Figure 1c). Similar to 2022, the success rate of projects in the feasibility study or concept stage announced for 2023 will be 1% at best. In summary, although likely not as wide as in 2022, the implementation gap is set to persist in 2023.

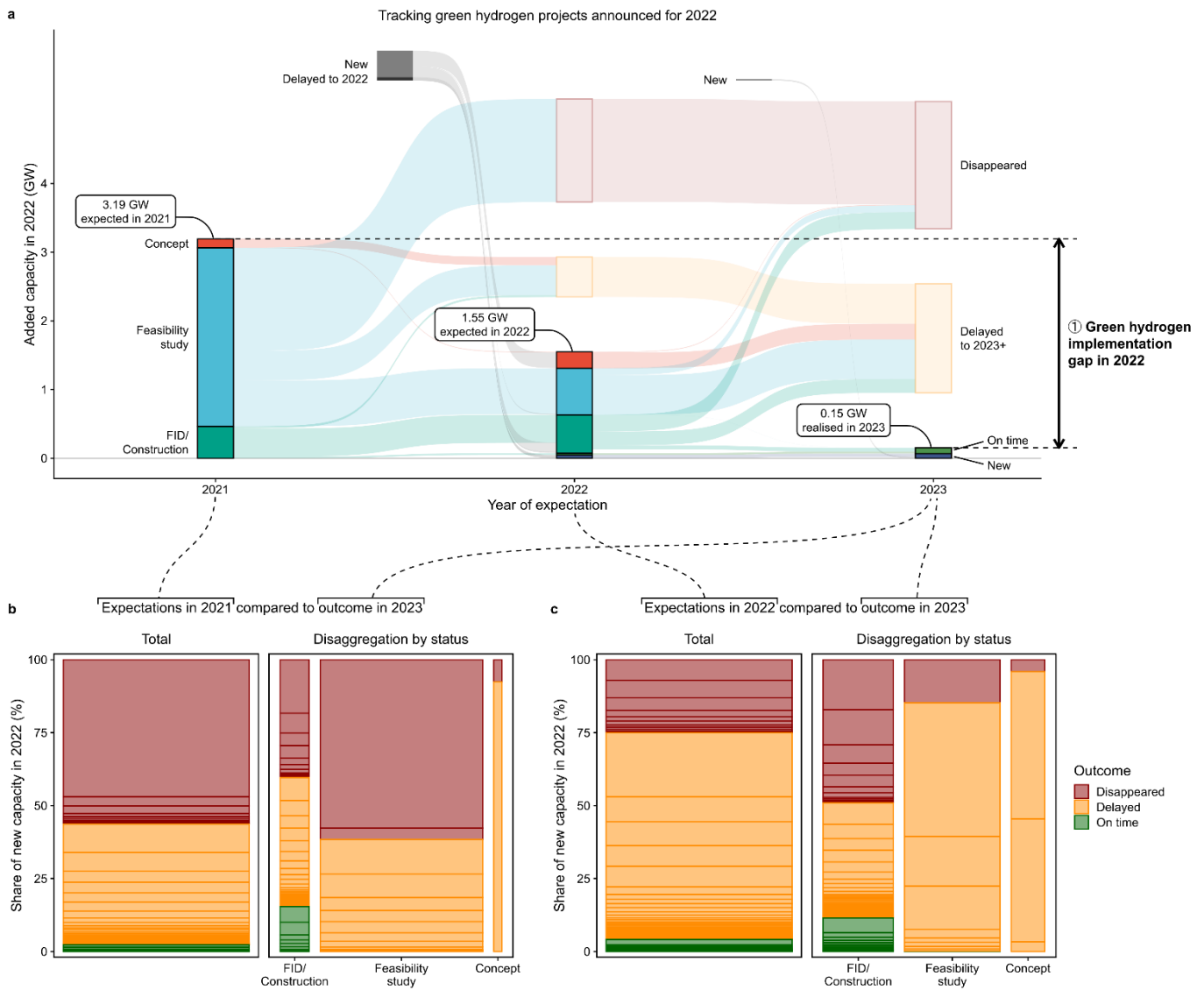

*Figure 3: The 2022 green hydrogen implementation gap. **a**,* Sankey diagram showing the development of green hydrogen projects announced for 2022 in terms of added electrolysis capacity. In 2021, added capacity in 2022 was expected to be 3.19 GW. This was revised downward to 1.55 GW in 2022. In 2023, it became clear that only 0.15 GW of new capacity had been installed in 2022. This results in a green hydrogen implementation gap of more than 3 GW in 2022. ***b-c**,* Percentage rates of success, delay and disappearance of green hydrogen projects announced to launch in 2022, comparing expectations in 2021 with the final outcome (b), and comparing expectations in 2022 with the final outcome (c). In both b and c, the left panel shows the total share, while the right panel shows the disaggregation by status. The width of the bars in the right panel corresponds to the share of total capacity (also compare with a). Within each colour, individual projects are shown as segments, ordered by size.

Although we do not investigate the individual factors contributing to project delay and disappearance, three main reasons for the poor performance of green hydrogen projects stand out. First, cost estimates for electrolysers have surged recently due to rising equipment and financial costs following the recent inflation[1]. Emerging evidence suggests that only the mass-producible electrolyser stack is set for rapid cost reductions due to standardisation, economies of scale and automated production[17]. However, other components of the electrolysis system such as balance of plant, engineering and construction[22], which constitute the largest cost share already today, are characterised by greater design complexity and require more customisation[17], making them less amenable to cost reductions[23]. Second, analysts have observed a lack of offtake agreements[18], resulting in substantial overcapacity of electrolyser manufacturing capacity. This may be explained by a limited willingness-to-pay for costly green hydrogen. Switching to hydrogen typically entails end-use investments that are often difficult to reverse (e.g. transforming steel production from a blast furnace to a direct reduction route, or switching from a diesel truck to a fuel-cell electric truck), carrying the risk of locking into an expensive and potentially scarce energy carrier or feedstock. Third, bridging the substantial cost gap and reducing investment risks requires hydrogen-specific support policies and regulation, even for countries with ambitious carbon pricing[24]. However, lagging implementation of support policies[1] and regulatory uncertainty regarding green hydrogen production standards in the EU and the US, although crucial to ensure climate benefits[25,26], have hampered growth.

What implications does the sobering track record of past project announcements have for the future of green hydrogen in ambitious climate change mitigation scenarios? To explore these ramifications, next we focus on the mid-term horizon towards 2030. First, we provide an overview of electrolysis requirements in 1.5°C scenarios, introducing the 2030 green hydrogen ambition gap. Second, we analyse the economic viability of surging project announcements and estimate subsidy volumes that would be required to realise all projects, leading to the 2030 green hydrogen implementation gap.

## The closing 2030 green hydrogen ambition gap

Comparing green hydrogen project announcements and requirements in 1.5°C scenario, we find that the green hydrogen ambition gap in 2030 has been gradually closing over the past three years, although this remains subject to substantial uncertainty (Figure 4). The closing ambition gap is the result of (i) a steadily growing project pipeline, and (ii) a decreasing role of green hydrogen in scenarios that have consistently reported electrolysis requirements in the past three years.

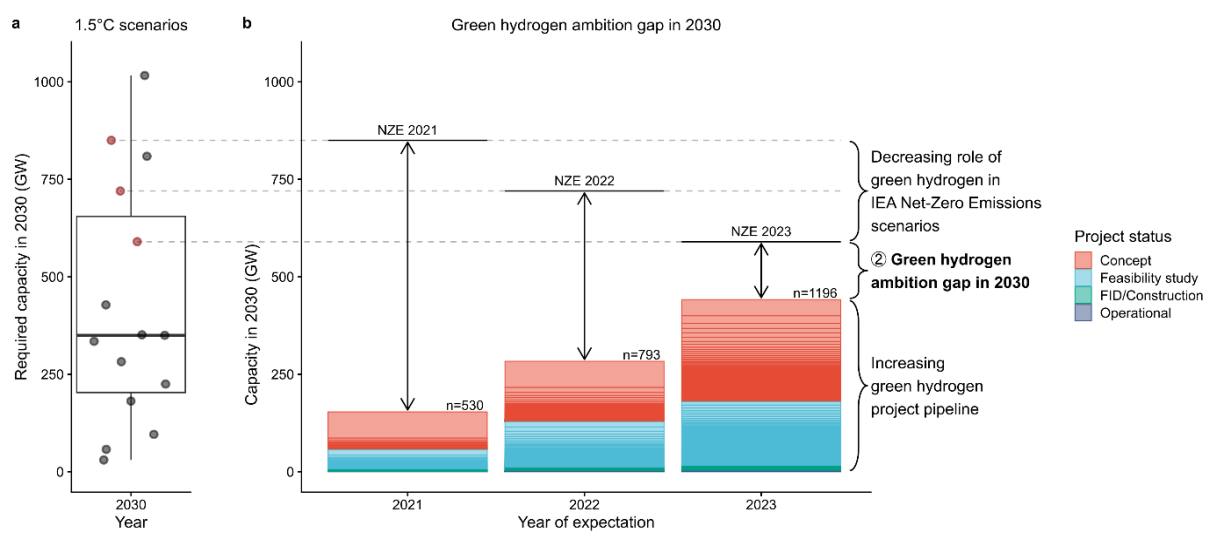

*Figure 4: The closing green hydrogen ambition gap in 2030. a,* Electrolysis capacity requirements for 2030 in 1.5°C scenarios (n=15), excluding one outlier scenario with 1700 GW in 2030 (see Extended Data Figure 2 and Methods). Whiskers indicate the range of scenarios from 30 – 1016 GW, underlining the high uncertainty around mid-term green hydrogen deployment. The box indicates the upper and lower quartile, spanning the interquartile range (203 – 655 GW). The horizontal line inside the box indicates the median at 350 GW. Each dot represents one scenario. Red dots indicate the IEA Net-Zero Emissions by 2050 (NZE) scenarios, which are the focus of panel b. *b,* Electrolysis capacity requirements in the IEA NZE scenarios and the project pipeline for 2030 by year of expectation. Of all scenarios shown in panel a, only the NZE scenarios provide annually updated electrolysis capacity in 2030 over the past three years, allowing us to analyse the changing role of green hydrogen in 1.5°C scenarios over time. The x-axis shows the year of expectation, corresponding to the year of publication of both the NZE scenarios and the Hydrogen Projects Database (see Methods). Individual projects are shown as segments within the coloured bars. Two trends emerge. First, the project pipeline for 2030 has increased three-fold in the past three years. Second, the NZE scenarios show a decreasing role of green hydrogen by 2030 in the past three years. Together, these two trends have been gradually closing the green hydrogen ambition gap in 2030, which we define as the difference between scenario requirements and project announcements. However, more than 97% of announced project capacity in 2030 are not yet backed by a final investment decision (FID).

Required levels of green hydrogen vary widely between different 1.5°C scenarios, in line with previous research[27] (Figure 4a). For 2030, this lack of consensus leads to an enormous range of 30 – 1016 GW of electrolysis capacity (excluding one outlier with 1700 GW), with a median of 350 GW and an interquartile range of 203 – 655 GW. This heterogeneity is the result of two key uncertainties. First, the upscaling of the nascent green hydrogen value chain from supply via infrastructure to demand is characterised by high uncertainty[28], which is exacerbated by the observation that project announcements have so far been a poor indicator of growth. This makes it difficult to carve out ambitious, yet feasible, mid-term pathways for the market ramp-up of green hydrogen. Second, although evidence shows that hydrogen and hydrogen-based electrofuels are promising for decarbonising end-use applications such as maritime shipping[29], aviation[30], and steel[31], the competition with alternative mitigation options such as direct electrification, biofuels or carbon capture and storage has not been fully settled and remains subject to uncertainty[32–34]. This structural uncertainty about competing mitigation options also persists in the long run and explains the high heterogeneity until 2050 (Extended Data Figure 2).

However, despite the high heterogeneity of green hydrogen in 1.5°C scenarios, we can identify an important trend in a subset of the scenarios: The IEA Net Zero Emissions by 2050 (NZE) scenarios, which have consistently reported values over the past three years[35–37], indicate a steady downward revision of required electrolysis in 2030 (Figure 4b). This is driven by an adjustment to reality following recent setbacks of green hydrogen as well as by fast progress of competing mitigation options, particularly a deep electrification of both passenger and truck transport as well as industrial and residential heat[37]. At the same time, the 2030 green hydrogen project pipeline has almost tripled from 153 GW to 441 GW, surpassing the requirements for 1.5°C in 10 out of 15 scenarios (Figure 4a-b). Thus, the green hydrogen ambition gap in 2030 has already closed for two thirds of the assessed scenarios, while it will likely also close for the NZE scenarios in 2024.

Although the convergence of the project announcements and scenario requirements is encouraging, the past green hydrogen implementation gap in 2022 and 2023 casts doubt on the reliability of ever-increasing project announcements. Of the 441 GW announced by 2030, 97% are still in the concept or feasibility study phase, which have exhibited critically insufficient success rates of 0% in 2022 and 1% in 2023 (see previous section). Achieving the level of ambition required in 1.5°C scenarios hinges on overcoming these high failure rates. Yet, how much policy support would be required to realise all project announcements?

## Estimating the 2030 green hydrogen implementation gap

The flipside of the closing green hydrogen ambition gap is the widening future green hydrogen implementation gap in 2030, which we define as the difference between project announcements and projects that are supported by policies. In this section we take green hydrogen project announcements for granted and estimate how much policy support would be required to realise all 441 GW by 2030. A crucial prerequisite for project realisation is cost parity with fossil fuels – but which fossil fuels exactly?

Natural gas is the most important competitor for green hydrogen across various end-use applications. This is illustrated by the IEA NZE scenario, where hydrogen replaces natural gas and natural gas-based grey hydrogen in roughly 90% of all hydrogen applications in 2030[37]. In industry, hydrogen competes with natural gas as a feedstock for ammonia production, as a reduction agent in steel production once a steel plant has switched to a direct reduction route, and as a fuel for process heat[31]. Within the power sector, hydrogen can provide dispatchable power and long-term energy storage in future highly renewable power systems[8] and therefore directly competes with future natural gas power plants with carbon capture and storage. Hydrogen also competes with natural gas in terms of infrastructure as many countries are currently envisaging to repurpose parts of their existing gas grid to transport hydrogen[38]. In addition, hydrogen competes with petroleum oil products in road transport, but faces stiff competition with electric vehicles[39] and increasingly also electric trucks[40,41]. In maritime shipping[29] and aviation[30], hydrogen-based electrofuels are important mitigation options, particularly when the potential for fossil carbon capture and utilisation (CCU) and direct-air carbon capture and storage (DACCS) is limited[33]. However, these electrofuels entail challenges[42] and will only become relevant at scale as the world approaches climate neutrality, a horizon beyond our mid-term focus until 2030. In the buildings sector, cost-optimal scenarios see no role for hydrogen in providing low-temperature heating as heat pumps are more efficient and hence more economical[43,44]. In summary, natural gas is the most important competitor in end-use applications, where hydrogen is expected to play a substantial role already in the next decade. Therefore, we use natural gas as a proxy for the competition between green hydrogen and fossil fuels (see Table 1 for parameters) Note that we compare natural gas and hydrogen based on their lower heating value, without distinguishing different end-use applications (see Limitations in the Methods).

*Table 1: Parameters for estimating the cost gap between green hydrogen and natural gas. The resulting cost gap for the central estimate is shown in Figure 5a-b. The scenario range column displays parameter values covered by the sensitivity scenarios (progressive and conservative, see Extended Data Figure 6, Extended Data Figure 7 and Methods). An empty cell in the scenario range column indicates that the central estimate also applies for the sensitivity scenarios. We use a payback period of 15 years, which is often shorter than the expected electrolyser lifetime, but represents the typical length of implemented policy support and is therefore more relevant for investment decisions. Displayed numbers are rounded. All currency values are in 2023US$.*

| Energy carrier | Parameter | Central estimate | Scenario range | Unit | Source/Comment |
|---|---|---|---|---|---|
| **Green hydrogen** | Electrolyser investment costs | 2023: 1850<br>2024 – 2030: Tech. learning driven by project announcements<br>2030 – 2045: Tech. learning driven by median 1.5°C scenario | 1700 – 2000 | $/kW(el) | IEA (2023)[1], see Methods for details of technological learning |
| | Stack share of electrolyser investment costs | 2023: 25<br>From 2024: Tech. learning | 14 – 29 | % | Ramboll (2023)[17] |
| | Learning rate: Stack | 18 | 15 – 20 | % | IEA (2023)[1] |
| | Learning rate: Balance of plant | 10 | 5 – 12 | % | IEA (2023)[1] |
| | Stack lifetime | 2024: 10<br>From 2030: 15 | 10 – 15 | yr | EPRI (2022)[45], IRENA (2020)[46] |
| | Electrolyser payback period | 15 | | yr | Hydrogen Europe (2023)[47], typical value for policy support such as auctions |
| | Full-load hours | 3750 | 3250 – 4250 | h/yr | Zeyen et al. (2024)[25], hourly matching with renewables |
| | Cost of capital | 8 | 6 – 10 | % | IEA (2023)[1] |
| | Electrolyser efficiency | 2024: 69<br>2045: 76 | | % | IEA (2023)[1], increase of approx. 0.3 pp per year |
| | Fixed operation and maintenance costs | 3 | 1.5 – 5 | % | Agora Industry and Umlaut (2023)[22] |
| | Transport and storage costs | 20 | | $/MWh | Generic value based on refs[1,24,48] (see Methods) |
| | Electricity price | 2024: 60<br>2030: 50<br>2045: 35 | 2024: 49 – 104<br>2030: 35 – 85<br>2045: 22 – 55 | $/MWh | Ueckerdt et al. (2024)[24], not only covering renewable LCOE, but also system costs (e.g. grid costs, see Methods) |
| **Natural gas** | Gas price | 2024: 19<br>2030: 22 | | $/MWh | Average of EU TTF and US Henry Hub spot market and future prices (see Methods) |
| | $CO_2$ price | **Scenario 1**: No $CO_2$ price<br>**Scenario 2**: EU climate target compatible (2024: 117, 2030: 149, 2035: 192, 2040: 246, 2045: 316) | **Scenario 2**: ±20% of central estimate | $/t$CO_2$ | Sitarz et al. (2024)[49] (see Methods) |
| | Emission intensity | 0.265 | | t$CO_2$/MWh | Ueckerdt et al. (2024)[24], incl. upstream emissions |

The competitiveness analysis reveals a substantial and prolonged cost gap between green hydrogen and natural gas (Figure 5a-b), defined as the difference between the levelised cost of hydrogen (LCOH) and the price of natural gas (see Methods and Table 1). Without carbon pricing, the cost gap of 160 $/MWh in 2024 implies that green hydrogen is initially more than 8 times as expensive as natural gas. By 2030, this cost gap decreases to 107 $/MWh, but persists at 70 $/MWh in 2045 (Figure 5a). Under an ambitious carbon price pathway in line with EU climate targets[49], the cost gap of 129 $/MWh in 2024 is narrower, but green hydrogen is still 3.5 times as expensive as natural gas. However, this cost gap steadily diminishes to 68 $/MWh by 2030, reaching cost parity in 2043 (Figure 5b). Thus, in the absence of carbon pricing, green hydrogen is expected to remain more expensive than its main competitor natural gas until well after 2045. Even with ambitious carbon pricing, the cost gap persists for almost two decades in our central estimate. Sustained support policies are therefore inevitable to foster green hydrogen growth.

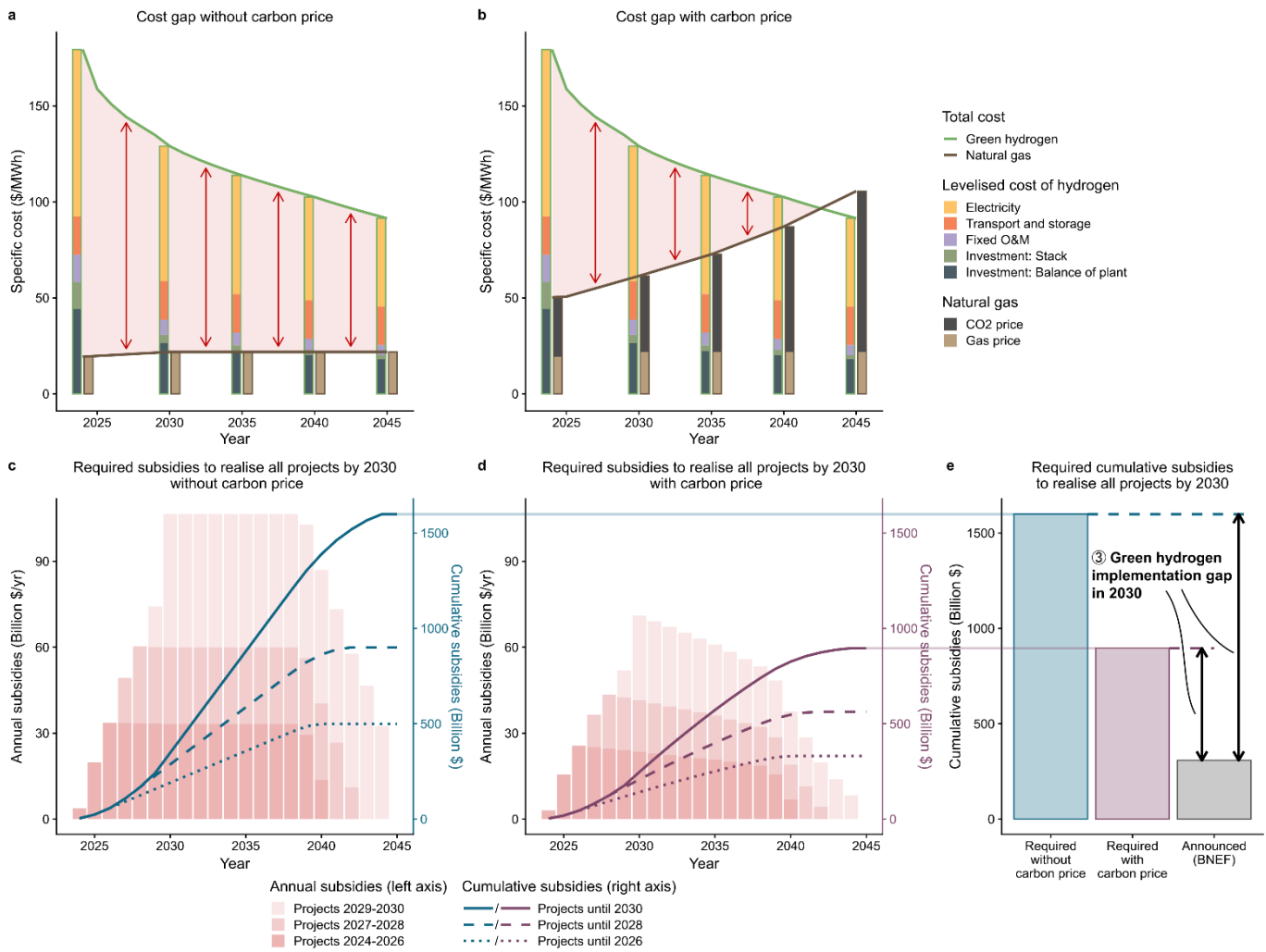

*Figure 5: The green hydrogen implementation gap in 2030.* **a-b,** Cost gap between the levelised cost of green hydrogen (LCOH) and the price of natural gas without carbon pricing (a) and with carbon pricing (b) for the central parameter estimate (see Methods). The red double-headed arrows and the light red shading indicate the cost gap that needs to be bridged by subsidies. The stacked bars indicate the decomposition of the LCOH and the total cost of natural gas for selected years (2024, 2030, 2035, 2040, 2045). For better visibility, the LCOH bar is slightly shifted to the left, and the natural gas bar to the right. Our 2030 LCOH estimates are in line with recent studies (see Extended Data Figure 5). **c-d,** Required subsidies to bridge the cost gap without carbon pricing (c) and with carbon pricing (d) in order to realise all project announcements until 2030 on time (see Methods). The bars show required annual subsidies (left axis), while the lines show required cumulative subsidies (right axis). **e,** Required cumulative subsidies to realise all projects announced until 2030 (without and with carbon pricing) in comparison to globally announced hydrogen subsidies as of September 2023 from BloombergNEF (BNEF)[50]. When calculating subsidies, we take currently implemented demand-side policies, which reduce pressure on supply-side subsidies, into account (see Methods and Extended Data Figure 3). Without carbon pricing, $1.6 trillion of subsidies are required to realise all projects announced until 2030. We explore the impact of more progressive and more conservative parameters for green hydrogen in Extended Data Figure 6 and Extended Data Figure 7, respectively (see column scenario range in Table 1). Note that c-e only show subsidies required for green hydrogen project announcements until 2030. Staying on a 1.5°C scenario requires substantial further subsidies after 2030, which, without carbon pricing, would need to be permanent (see Extended Data Figure 8).

The main contributors to the LCOH are electricity costs, electrolyser investment, and transport and storage costs (Figure 5a-b). Although investment costs of electrolysers have recently surged due to higher equipment and financing costs[1,17], this trend is expected to reverse soon due to learning-by-doing and economies of scale. Note again, that in order to estimate the volume of required subsidies we consider a scenario where all project announcements until 2030 are realised on time, while after 2030 cost reductions are driven by the median electrolysis capacity in 1.5°C scenarios (see Methods and Extended Data Figure 3). As a consequence, electrolyser investment costs fall quickly, especially for the electrolyser stack, which can be mass-produced in automated factories[17], while the balance of plant requires more customisation and is thus less amenable to rapid cost reductions[23]. In our central estimate, electrolyser investment costs drop from 1850 $/kW in 2023, to 701 $/kW in 2030, and to 493 $/kW in 2045 (Extended Data Figure 4). We calculate the LCOH with a payback period of 15 years. Although this is shorter than the expected electrolyser lifetime, it represents the typical length of implemented policy support such as auctions[47] and is therefore more

relevant for investment decisions than the technical lifetime. Our 2030 LCOHs are consistent with recent studies (Extended Data Figure 5).

The cost gap between green hydrogen and natural gas is robust across a wide range of progressive and conservative parameter values, especially in scenarios without carbon pricing. In these scenarios, green hydrogen always remains more expensive than natural gas until after 2045, though the size of the cost gap varies (Extended Data Figure 6a, Extended Data Figure 7a). In scenarios with carbon pricing, cost parity is eventually achieved, but the timing is uncertain. With progressive parameters, cost parity could be reached as early as 2034 (Extended Data Figure 6b), whereas with conservative parameters, this may not occur until well after 2045 (Extended Data Figure 7b). This underlines the considerable uncertainty about when green hydrogen will become cost-competitive solely due to carbon pricing, highlighting the need for suitable policy instruments to reduce investment risks.

Bridging the cost gap between green hydrogen and natural gas requires substantial subsidies, or equivalent regulation, for several decades. To recover their costs, green hydrogen projects must sell hydrogen at their respective LCOH throughout the payback period. Assuming that the willingness-to-pay for green hydrogen matches that of natural gas, the cost gap determines the specific per-MWh hydrogen subsidy a project requires. Hydrogen suppliers could secure such policy support through pay-as-bid reverse auctions for a fixed premium paid upon delivery. In order to estimate annually required subsidies, we multiply the specific hydrogen subsidies with the announced new green hydrogen capacity (Extended Data Figure 3a). We also account for currently implemented demand-side policies that support green hydrogen capacity without requiring additional subsidies (see Methods).

Annual subsidies required to realise all projects by 2030 are bell-shaped, which results from the payback period, with the height and timing of the peak varying by scenario (Figure 5c-d, left axis). Without carbon pricing, required annual subsidies rise sharply to a plateau of more than $100 billion per year throughout the 2030s (Figure 5c). With carbon pricing, required annual subsidies peak at $71 billion per year in 2030 (Figure 5d). The resulting cumulative subsidies to realise all 441 GW of green hydrogen projects by 2030 are S-shaped (Figure 5c-d, right axis). Without carbon pricing, required cumulative subsidies amount to $1.6 trillion ($1.2 – 2.6 trillion range, Extended Data Figure 6c, Extended Data Figure 7c). With carbon pricing, required cumulative subsidies are still $0.9 trillion ($0.3 – 2.1 trillion range, Extended Data Figure 6d, Extended Data Figure 7d). While already substantial, these figures only pertain to the 2030 project pipeline. Aligning green hydrogen with 1.5°C scenarios after 2030, would require considerably higher subsidies. Without carbon pricing, the cost gap does not close by 2045, leading to cumulative subsidies of $6.3 trillion by 2045 ($4.4 – 10.8 trillion range, Extended Data Figure 8), further rising afterwards. This shows that permanently subsidising green hydrogen against cheap fossil fuels will likely be prohibitively expensive in the long term, highlighting the key role of carbon pricing in closing the implementation gap.

Due to a substantial discrepancy between required and announced subsidies, a wide 2030 green hydrogen implementation gap arises (Figure 5e). Cumulative subsidies required to realise all project announcements by 2030 exceed currently announced subsidies, estimated at $308 billion as of September 2023[50], by a factor of three with carbon pricing and five without. There are counteracting uncertainties regarding this estimate, as announced subsidies are likely to increase in the future, but challenges may arise during their implementation (see Methods). Even if all currently announced global subsidies were immediately available, without carbon pricing this would only support 56 GW by 2030 (34 – 72 GW range, 8 – 16 % of the project pipeline, see Extended Data Figure 9). This is outweighed by the impact of already-implemented demand-side policies in almost all scenarios, underlining the importance of demand-side regulation for fostering green hydrogen growth.

## Discussion and conclusion

The past and future of green hydrogen is characterised by three gaps, reflecting the challenges of scaling-up a novel and yet uncompetitive energy carrier that requires dedicated policy support. First, the 2022 implementation gap shows that only 2% of announced green hydrogen capacity was realised on time. Second, the 2030 ambition gap has gradually closed, as the project pipeline and requirements in 1.5°C scenarios increasingly converge. Third however, this has led to a wide 2030 implementation gap as enormous subsidies would be required to realise all projects by 2030 – and even more to put green hydrogen on track for 1.5°C in the long term.

The very high past failure rates indicate a limited reliability of project announcements published by industry, especially for projects without a final investment decision. Although sobering, this can provide valuable insights for realistic scale-up analyses of hydrogen[28] and other low-carbon energy technologies in the feasibility literature[51–54], especially for analyses that use uncertain project announcements as a data input[55,56]. System planners, policymakers and society should interpret the increasingly steep growth suggested by recent project announcements (refer to Figure 1) with caution, while focusing on scale-up challenges such as lacking competitiveness and the need for policy support.

To close the green hydrogen implementation gap, policymakers need to bridge the cost gap to fossil fuels and de-risk hydrogen investments. This requires a balanced policy mix and a robust strategy to navigate the following key uncertainties and risks:

1. **Prolonged green hydrogen scarcity.** The huge past and future implementation gaps indicate that green hydrogen will likely fall short of the requirements in 1.5°C pathways. Even if policy support is strengthened, it remains uncertain whether this is sufficient to drive the necessary hydrogen investments. Realising current project announcements would require unprecedented growth rates, exceeding even the fastest-growing energy technologies in history, solar photovoltaics. Given that green hydrogen technologies are more complex, less standardisable and require new infrastructure, all of which slow down technology diffusion[23], realising such unprecedented growth is unlikely.

2. **High policy costs.** Current hydrogen policy instruments often seek to spur hydrogen investments by bridging the cost gap to fossil fuels through supply-side subsidies such as fixed-premium auctions. However, as we have shown, this approach requires not only excessive subsidy volumes, but also strong perseverance as policy support could be required for several decades, or even permanently without carbon pricing or strong demand-side regulation. The narrative of kickstarting a "hydrogen economy" with only a short-term policy push, after which green hydrogen becomes cost-competitive and scales up on its own, is likely misleading and raises false hopes.

3. **Uncertain long-term role.** Hydrogen's primary role in climate change mitigation is to replace fossil fuels in hard-to-electrify sectors. However, strong political support for hydrogen is often accompanied by overconfidence in its potential[15], resulting in conflicting visions about hydrogen's future role. Many global climate change mitigation scenarios show a modest long-term share of hydrogen in final energy of 5-15%[2,37,57], focusing on key end-uses where hydrogen is highly valuable due to a lack of alternatives[5]. In stark contrast, incumbent actors from gas, heat, industry and transport tend to endorse a wide use of hydrogen across sectors[58], even in end-uses like residential heat, where electrification is cheaper, more efficient and readily available[37,2,57,43]. Uncertainties remain around hydrogen's role in complementing electrification of heavy transport and industrial heat[11,32,37].

Disregarding these uncertainties and risks, and instead focusing on supply-side subsidies with the expectation of abundant low-cost green hydrogen in the future, risks crowding out available and more economical options, thereby delaying climate change mitigation. To minimise these risks while safeguarding the scale-up of green hydrogen, we draw two key policy conclusions.

First, supply side subsidies, which reduce the investment risk of electrolysis projects, should be complemented by demand-side policies that guide hydrogen to its most valuable use cases by increasing their willingness-to-pay. The benefit of demand-side measures is illustrated by the European Hydrogen Bank's recent inaugural auction, which resulted in surprisingly low successful bids of 0.37 – 0.48 €/kg[59], compared to a similar auction in the UK, which only received high bids equivalent to 9.40 €/kg[60]. Aside from regional heterogeneity, this stark difference can be attributed to the EU's demand-side quotas, such as the mandatory 42% green hydrogen share in industry by 2030 under the Renewable Energy Directive III[61], and mandates for hydrogen-based electrofuels under ReFuelEU Aviation[62] and FuelEU maritime[63] regulations. Complementary demand-side policies can thus reduce the pressure on supply-side subsidies, helping to close the implementation gap.

Second, policymakers should plan the transition from subsidies to market mechanisms. In the short run, achieving strong near-term hydrogen growth is crucial to keep 1.5°C scenarios within reach. This requires strong policies such as subsidies to directly bridge the cost gap, minimise investment risks and initialise a hydrogen market. However, as hydrogen technologies and markets mature, policy support should shift to market-based mechanisms in order to (i) reduce policy costs, (ii) reveal the full hydrogen costs to markets and consumers, and (iii) create a level playing field with other mitigation options. The most-important technology-neutral strategy is ambitious carbon pricing. However, as carbon prices are currently too low and too uncertain in the future, complementary instruments are required to de-risk the remaining uncertainties. These include technology-neutral auctions of carbon contracts-for-difference[64], which hedge investors against unpredictable prices by covering the difference between emissions abatement costs and carbon prices, as well as tradable, technology-neutral quotas for e.g. low-carbon materials, fostering green lead markets.

In summary, a comprehensive policy strategy for green hydrogen should include targeted demand-side measures and a gradual transition from subsidies to market mechanisms. In the short term, this would de-risk early investment at manageable costs, guiding hydrogen to its most valuable use cases. In the long term, this would transfer investment risks and competition between hydrogen and other mitigation options to the market, thereby establishing a credible commitment for climate change mitigation while spurring green hydrogen growth.

# Methods

### Overview

Our approach is split into three parts. First, we track green hydrogen project announcements to quantify the green hydrogen implementation gap in 2022 and 2023. Second, we compare project announcements with 1.5°C scenarios to show the 2030 green hydrogen ambition gap. Third, we conduct a competitiveness analysis of green hydrogen and its main competitor natural gas in order to estimate required annual and cumulative subsidies, leading to the 2030 green hydrogen implementation gap.

### Green hydrogen project database

We use data of electrolysis project announcements from the IEA Hydrogen Production Projects and Infrastructure Database[65], incorporating three database snapshots from 2021, 2022 and 2023. We only include project announcements for electrolysers that include a year of project launch, have a meaningful status (not "Other" or "Other/Unknown"), and report a capacity value. We do not filter for the type of electricity as this is often unknown. These criteria lead to 611 projects in the 2021 snapshot, 879 projects in the 2022 snapshot, and 1296 projects in the 2023 snapshot. In the 2023 snapshot, the status categories "FID" and "Under Construction" were merged into a single category "FID/Construction". In order to ensure consistent status categories across all snapshots, we also merge these categories in the 2021 and 2022 snapshot. Projects with a "DEMO" status are allocated as "Operational", "FID/Construction" or "Decommissioned", depending on whether they are still running, announced for the future, or have been decommissioned, respectively. Confidential projects are distributed to all regions in proportion to the share of capacity from non-confidential projects, but cannot be tracked across database snapshots. Following the IEA's Creative Commons (CC) license, we note that this is a work derived by us from IEA material and we are solely liable and responsible for this derived work. The derived work is not endorsed by the IEA in any manner.

### Tracking green hydrogen projects

Each project has a unique reference number that stays the same across all database snapshots as confirmed by the IEA in personal correspondence. This enables us to track the development of project announcements over time (Figure 3 and Extended Data Figure 1). We account for changing capacity of projects between two database snapshots by adding dummy projects. However, for simplicity these are not explicitly shown, but remain visible as minor differences between stocks and flows in the Sankey diagrams. The reported rates of disappearance, delay and success (Figure 3b-c and Extended Data Figure 1b-c) only refer to projects as expected in 2021 and 2022, respectively.

### Green hydrogen in 1.5°C scenarios

As an indicator of green hydrogen requirements in stringent climate mitigation scenarios, we collect electrolysis capacity from a wide range of 1.5°C scenarios (see Extended Data Figure 2). Due to limited reporting of numerical data in text or tables, in some cases we resort to extracting data from graphics using WebPlotDigitizer, which has been shown to be reliable[66]. All data sources are clearly stated in the files provided on GitHub. If scenarios do not report electrolysis capacity, we convert production quantities into corresponding electrolysis capacity, assuming 3,750 full-load hours, 69% efficiency, and the lower heating value of hydrogen, 33.33 kWh/kg. Due to these approximations, reported electrolysis requirements in 1.5°C scenarios are associated with uncertainties. In Figure 4, we focus on the Net Zero Emissions by 2050 Scenarios of the IEA[35–37], which is the only scenario that has reported electrolysis capacity annually in the past three years and therefore enables us to track the closing green hydrogen ambition gap in 2030 over time.

### Levelised cost of hydrogen

To gauge the economic viability of steeply increasing project announcements, we conduct a simplified competitiveness analysis between green hydrogen and its main competitor natural gas for three scenarios (central, progressive, conservative), each with and without an ambitious carbon price. All required parameters are shown in Table 1 and made available on GitHub. Some parameters change over time, in which case we use linear interpolation in between.

For green hydrogen, we calculate the levelised cost of hydrogen (LCOH) for each year from 2024, using the annuity method and broadly following the system boundaries outlined in ref[22], but adding generic transport and storage costs (see below). Omitting time indices, the LCOH is calculated as

$$\text{LCOH} = \frac{1}{\eta}\left((a(r,\tau) + FOM)\frac{I_{BOP}}{\text{FLH}} + (a(r,\tau_{stack}) + FOM)\frac{I_{stack}}{\text{FLH}} + p_{elec}\right) + \text{VOM} \quad (1)$$

where $\eta$ denotes the electrolyser efficiency, $a(r,\tau) = \frac{r}{1-(1+r)^{-\tau}}$ the annuity factor, $r$ the cost of capital, $\tau$ the payback period of the entire electrolyser in years (which can be shorter than the technical lifetime), $\tau_{stack}$ the lifetime of the electrolyser stack in years, FOM the fixed operation and maintenance costs as a percentage of the specific investment costs, $I_{BOP}$ the specific investment cost of the electrolyser's balance of plant (BOP) and other engineering work, $I_{stack}$ the specific investment cost of the electrolyser stack, FLH the full-load hours, $p_{elec}$ the electricity price, and VOM the variable operation and maintenance costs, which are generic transport and storage costs.

We use a generic value for these transport and storage costs (see Table 1), based on literature values. In the short term, we assume that storage costs for expensive steel tanks and battery storage dominate, while in the long term hydrogen could be traded over long distances such that costs for transport (0.5 $/kg for a 3000 km pipeline[1], equivalent to 15 $/MWh) and cavern storage (approximately 5 $/MWh[48]) dominate. While a recent study projects much higher levelised costs of storing hydrogen in salt caverns, equivalent to 22 – 58 $/MWh[67], these could be mitigated as not all hydrogen will need to be stored. Overall and in agreement with recent studies[24], we summarise all transport and storage costs into a generic value of 20 $/MWh. Note that transport via maritime shipping or further distribution to refuelling stations for road transport would be costlier.

The electricity price paid by electrolysers highly depends on the specific supply case and the regulatory definition of green hydrogen with respect to spatio-temporal matching and additionality[25,26]. Flexible operation and a direct connection to a renewable energy source reduces the price as electrolysers can tap into hours where electricity is cheap and abundant. Grid-connected electrolysers need to pay grid fees on top of electricity prices, but can run at higher full-load hours. We account for these effects in an aggregated manner by using the same range of electricity prices as in ref[24].

We separate the total specific investments costs of the electrolyser, $I$, into the stack $I_{stack}$ and the balance of plant $I_{BOP}$. This has two reasons. First, the stack often needs to be replaced earlier than the rest of the electrolyser, which is why we distinguish two annuities in equation (1)[22]. Second, the stack is much more granular and therefore more susceptible to undergo cost improvements through technological learning[17]. Technological learning results from economies of scale and mass production and reduces specific investment costs of both $I_{BOP}$ and $I_{stack}$ in year $t$ according to

$$I_t = I_{2023} \left(\frac{C_t}{C_{2023}}\right)^{\log_2(1-LR)} \quad (2)$$

where $C_t$ denotes global cumulative electrolysis capacity in year t, $LR$ denotes the learning rate, defined as the percentage decrease of specific investment costs for each doubling of cumulative capacity. For the base year 2023, we assume that all remaining uncertain projects are built on time, resulting in $C_{2023} = 1.86\ GW$ (see Figure 1). Technological learning is driven by cumulative project announcements until 2030, and subsequently by the median 1.5°C scenario (see Extended Data Figure 3). We use different learning rates for the electrolyser stack and the BOP, leading to the investment costs shown in Extended Data Figure 4.

### Estimating required subsidies

We estimate annual and cumulative subsidies that would be required to realise all global green hydrogen project announcements by 2030 (Figure 5, Extended Data Figure 6 and Extended Data Figure 7), and to continue on a median 1.5°C pathway after 2030 (Extended Data Figure 8). To that end, we compare the LCOH with the price of natural gas, green hydrogen's main competitor (see main text). We estimate the price of natural gas as the average of the EU trading point Title Transfer Facility (TTF) in the Netherlands and the US trading point Henry Hub, using spot market prices in 2024 and future prices in 2030. As of May 2024, this leads to a gas price of 19 $/MWh in 2024, rising to 22 $/MWh in 2030, after which we assume it to remain constant. In the illustrative scenario with carbon pricing, we consider a $CO_2$ price pathway that is consistent with meeting the EU climate targets for the sectors covered by the EU Emissions Trading System (ETS) such as industry and energy supply[49]. The $CO_2$ cost per MWh of natural gas is the product of the emissions intensity, including upstream methane emissions[24], and the carbon price per ton of $CO_2$. We denote the total cost of natural gas as $p^{gas}$, which includes $CO_2$ costs if applicable. This leads to an instantaneous cost gap between green hydrogen and natural gas in year t of

$$\Delta p_t = LCOH_t - p_t^{gas} \quad (3)$$

which is depicted in Figure 5a-b, Extended Data Figure 6a-b, and Extended Data Figure 7a-b.

A green hydrogen project completed in year $t'$ must sell hydrogen at $\text{LCOH}_{t'}$ for the duration of payback period $\tau$ to recover its costs. Thus, In each year $t = [t', t' + \tau[$, the project requires subsidies to bridge the cost gap to the current gas price $p_t^{gas}$. Therefore, required annual subsidies accumulate over time due to projects built in previous years. For example, in 2026, projects that were built in 2024 face a cost gap of $LCOH_{2024} - p_{2026}^{gas}$, projects that were built in 2025 face a cost gap of $LCOH_{2025} - p_{2026}^{gas}$, and projects that were built in 2026 face a cost gap of $LCOH_{2026} - p_{2026}^{gas}$.

The cost gap has to be bridged for all new electrolysis capacity built in year $t'$, accounting for capacity that is already supported by demand-side policies, $\Delta C_{t'}$. With full-load hours FLH, the electrolyser efficiency $\eta$, and the payback period $\tau$, the required *annual* subsidies in year t are:

$$S_t^{annual} = \sum_{t'=\max\{2024, t-\tau\}}^{t} \Delta C_{t'} \cdot \text{FLH}_{t'} \cdot \eta_{t'} \cdot \max\{0, \text{LCOH}_{t'} - p_t^{gas}\} \quad (4)$$

Thus, realisation of green hydrogen projects built in $t'$ requires subsidy payments for the full payback period $t = [t', t' + \tau[$ as long as $LCOH_{t'} > p_t^{gas}$. We now explain how we obtain $\Delta C_t$.

We consider that demand-side policies such as end-use quotas increase the willingness-to-pay for hydrogen and thus reduce the remaining cost gap and required policy costs. We represent this mechanism by accounting for implemented demand-side measures, estimated at 7 Mt hydrogen per year in 2030 according to ref[1], which we translate into supported cumulative electrolysis capacity in 2030 using the lower heating value (LHV) of hydrogen, 33.33 kWh/kg, according to:

$$C_{2030}^{supported} = \frac{7\ \text{Mt/yr} \cdot LHV}{FLh \cdot \eta} \quad (5)$$

Distributing $C_{2030}^{supported}$ proportionally to yearly capacity additions within the time frame $t = [2024,2030]$ yields $\Delta C_t^{supported}$, with $C_{2030}^{supported} = \sum_{t=2024}^{2030} \Delta C_t^{supported}$ (see hatched bars in Extended Data Figure 3a). Thus, $\Delta C_t$ follows as

$$\Delta C_t = \Delta C_t^{projects} - \Delta C_t^{supported}, \quad (6)$$

where $\Delta C_t^{projects}$ denotes new projects announced to launch in year t. While we focus on subsidies required for project announcements until 2030, in Extended Data Figure 8 we extend our analysis beyond 2030 such that $\Delta C_t^{projects}$ represents annual capacity additions in line with the median 1.5°C scenario (see Extended Data Figure 3). Thus, equation (4) and (6) determine the annual subsidies, $S_t^{annual}$, required to build $\Delta C_t^{projects}$, given that $\Delta C_t^{supported}$ are already supported by demand-side regulation.

Correspondingly, the required *cumulative* subsidies payable until t are:

$$S_t^{cumulative} = \sum_{t'=2024}^{t} S_{t'}^{annual} \quad (7)$$

To answer how many subsidies will be required in total over the full payback period of a project, we have to look $\tau$ years beyond the project launch date. In Figure 5c-e, Extended Data Figure 6c-e, and Extended Data Figure 7c-e, we therefore focus on cumulative subsidies required for projects announced until 2030, which due to a payback period of 15 years imply subsidy payments until 2045.

Limitations

The data quality of the IEA Hydrogen Production and Infrastructure Projects Database[65] may be limited. This is particularly important for the project tracking, where single project announcements can have a large influence on the rates of project disappearance, delay and success (see Figure 3 and Extended Data Figure 1). Due to the high number of project announcements, we did not conduct additional market research and therefore also cannot correct inaccurate project data. Furthermore, by their nature, confidential projects cannot be tracked across database snapshots.

The data quality of the electrolysis requirements in 1.5°C scenarios is limited due to heterogenous sources and limited numerical reporting of scenario data accompanying the reports. In several cases we have to infer electrolysis capacity from green hydrogen production values. Thus, Figure 4 and Extended Data Figure 2 only show publicly available data and should not be interpreted as numerically exact.

Estimating subsidies requires several simplifications to reduce complexity. First, we do not distinguish between regions or end-use applications when calculating the cost gap between green hydrogen and its fossil competitors. This would substantially increase complexity as we would need to represent different end-use applications and assume their region-specific uptake and shares in time. Instead we use natural gas as the main competitor, which is representative for most use cases of green hydrogen in industry and energy supply. The cost comparison is based on the lower heating value, which is a valid assumption for industrial process heat or power generation, while for ammonia or DRI (direct reduced iron) production, green hydrogen can be used slightly more efficiently than natural gas. We also neglect additional costs associated with hydrogen on the end-use side. This simplification has a low impact on our results as additional costs are typically low or even zero. Some applications can simply substitute grey with green hydrogen with no additional costs (e.g. ammonia production), while additional investment costs in other applications are low compared with fossil applications (e.g. DRI-based steel plants or hydrogen boilers).

Second, concerning specific investment costs, we do not include a feedback from policy-supported deployment and associated technological learning, but instead assume that all project announcements are built on time. Third, we use the annuity method instead of a discounted cash flow analysis to calculate LCOHs, neglecting the impact of future changes in cost components, which is particularly relevant for electricity. However, many green hydrogen projects will require new dedicated renewable energy plants or long-term contracted power-purchase agreements, making constant electricity prices over the payback period a valid simplification. Fourth, we of course have to carefully choose values for all parameters presented in Table 1, which inevitably involves uncertainties. To analyse the impact of these uncertainties, we conduct two sensitivity analyses that explore a wide range of progressive and conservative parameter settings (see Extended Data Figure 6 and Extended Data Figure 7). We also compare our LCOH in 2030 to recent literature values (Extended Data Figure 5). Fifth, we do not consider the option that green hydrogen projects could pay back a part of received subsidies once they are profitable relative to natural gas in the future because (i) this would require a contract-for-differences that allows for this option and (ii) most projects do not become competitive within the time horizon we analyse. Sixth, we do not include factors other than costs that influence the realisation of a project. In reality, project realisation depends on a plethora of highly context-specific decision factors beyond pure economic viability, which is beyond the scope of this work. Seventh, we assume that demand-side measures translate into electrolysis capacity without requirements for additional subsidies. Additional demand-side measures such as an increase in end-use hydrogen quotas could further reduce the need for subsidies and thus decrease the implementation gap in the future. For simplicity, we also assume that all demand from end-use quotas is supplied by green hydrogen.

Lastly, the data quality of global announced hydrogen subsidies from BloombergNEF (BNEF) may be limited. The subsidy estimate for the United States is particularly uncertain as the production tax credits of the Inflation Reduction Act[12] are uncapped. Therefore, BNEF estimates US subsidy announcements based on project announcements, which implies uncertainties. Furthermore, tracked subsidies cover not only green hydrogen, but also other sources of low-carbon hydrogen, which we optimistically compare to subsidy requirements required only for green hydrogen project announcements. The global subsidy volume of $308 billion for low-carbon hydrogen as of September 2023 therefore only serves as a snapshot. While we acknowledge that this value will soon be outdated, it still provides a useful indication. Nevertheless, it should be interpreted with caution as implemented subsidies critically depend on governments' future commitment to foster the hydrogen market ramp-up.

## Data availability

All data is publicly available on GitHub (see Code availability). This includes the IEA Hydrogen Production and Infrastructure Projects Database from 2021 to 2023[65], electrolysis requirements in 1.5°C scenarios from various sources, techno-economic data for the competitiveness analysis (see Table 1), data of announced hydrogen subsidies by BloombergNEF[50], and LCOH values in recent studies. All data files include a column that indicates the original source.

## Code availability

The R model code, including all data, to perform the analyses and reproduce all figures is available on GitHub:

https://github.com/aodenweller/green-hydrogen-gap

## Acknowledgements

We gratefully acknowledge funding from the Kopernikus-Projekt Ariadne by the German Federal Ministry of Education and Research (grant nos. 03SFK5A, 03SFK5A0-2, A.O., F.U.) and the HyValue project (grant no. 333151, F.U.). We thank the IEA, particularly J.M. Bermudez, for providing the Hydrogen Projects Database and for answering related questions. We thank Robert Pietzcker for brainstorming about figures, Philipp Verpoort for cost parameters, and Joanna Sitarz for carbon price data.

## Author Contributions

A.O. suggested the initial research question, which F.U. extended. A.O. and F.U. collected cost data. A.O. collected 1.5°C scenario and LCOH data. A.O. performed the analyses and created the figures. A.O. and F.U. interpreted the results. A.O. wrote the manuscript with contributions from F.U.

## Competing interests

The authors declare no competing interests.

# Extended Data Figures

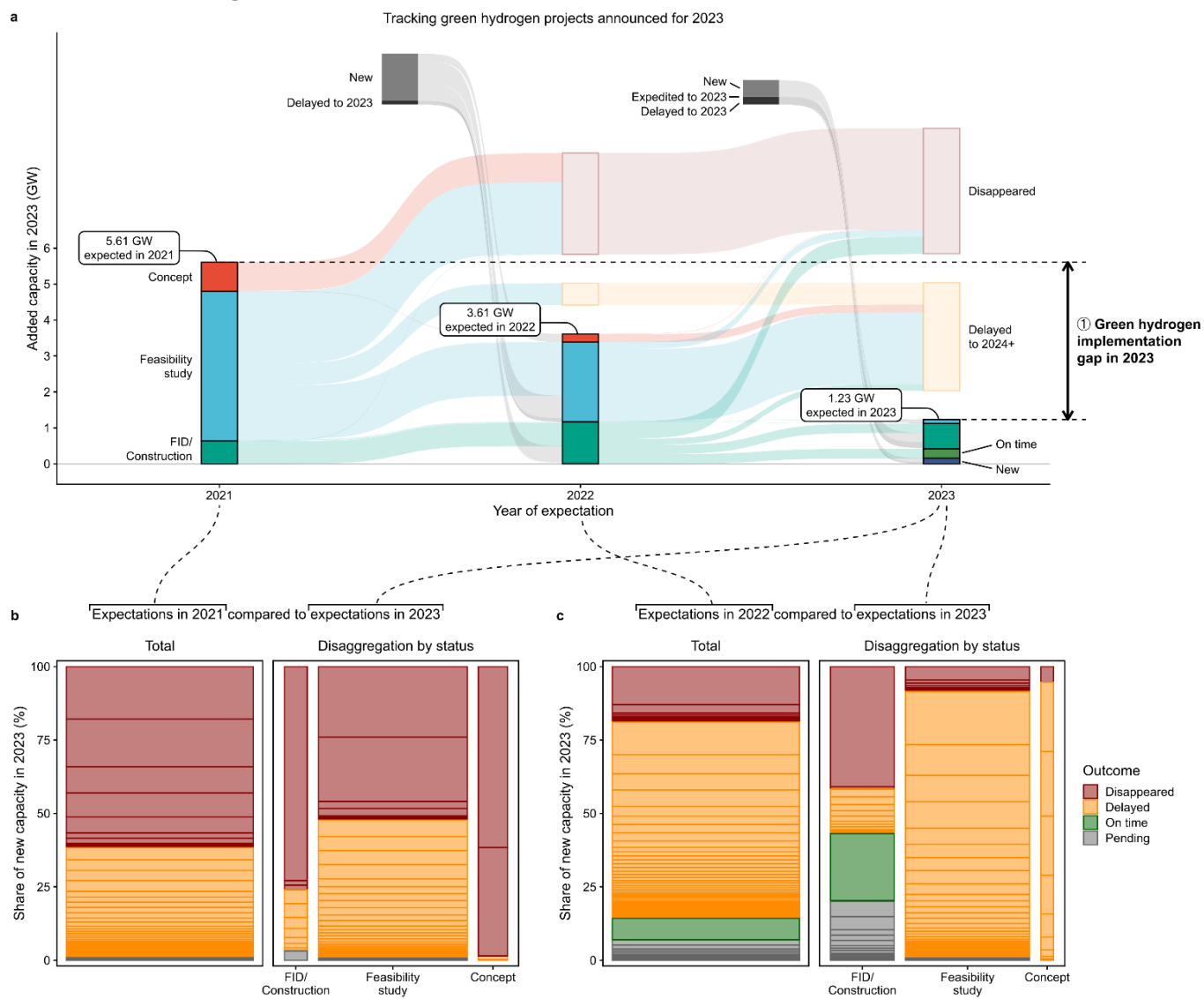

***Extended Data Figure 1: The 2023 green hydrogen implementation gap.*** *Analogous to Figure 3 for the year 2023.* ***a,*** *Sankey diagram showing the development of green hydrogen projects announced for 2023 in terms of added electrolysis capacity. The most recent database version is from October 2023 such that the final added capacity in 2023 is still uncertain.* ***b-c,*** *Percentage rates of success, delay and disappearance of green hydrogen projects announced to launch in 2023, comparing expectations in 2021 with expectations in 2023 (b), and comparing expectations in 2022 with expectations in 2023 (c). In contrast to Figure 3, the outcome of some project announcements for 2023 is still pending.* ***b,*** *Comparing expectations in 2021 with expectations in 2023.* ***c,*** *Comparing expectations in 2022 with expectations in 2023. In both b and c, the left panel shows the total share, while the right panel shows the disaggregation by status. The width of the bars in the right panel corresponds to the share of total capacity (also compare with a). Within each colour, individual projects are shown as segments, ordered by size. Note that the time difference between the first database snapshot in 2021 and 2023 is longer than in Figure 3, which focused on 2022. To ensure that the comparison is based on the same lead time between expectations and outcome, Figure 3b should therefore be compared with Extended Data Figure 1c. While some uncertainty remains, no more than 15% of expected capacity will be realised in 2023 and the success rate of projects in the feasibility study will be 1% at best (c).*

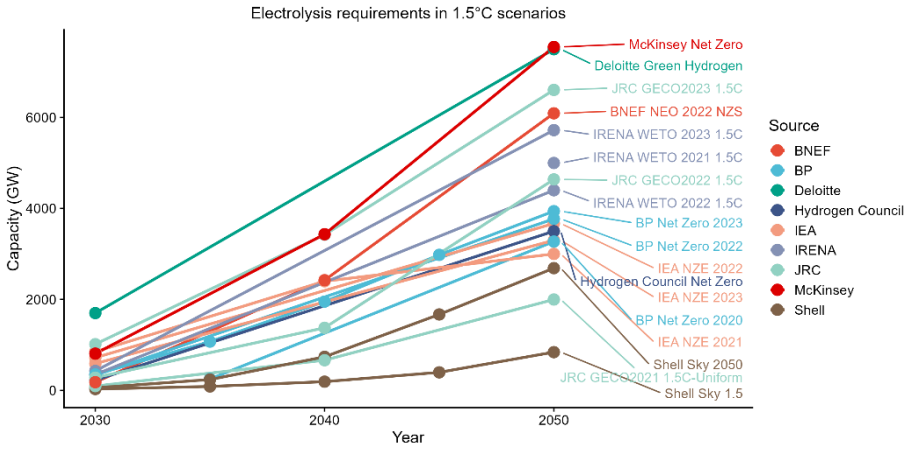

*Extended Data Figure 2: Electrolysis requirements in 1.5°C scenarios from 2030 to 2050.* The assessed scenarios show a wide range, underlining the high uncertainty surrounding the green hydrogen market ramp-up. If scenarios do not report requirements in terms of electrolysis capacity, we convert production quantities into corresponding electrolysis capacity, which implies uncertainties (see Methods).

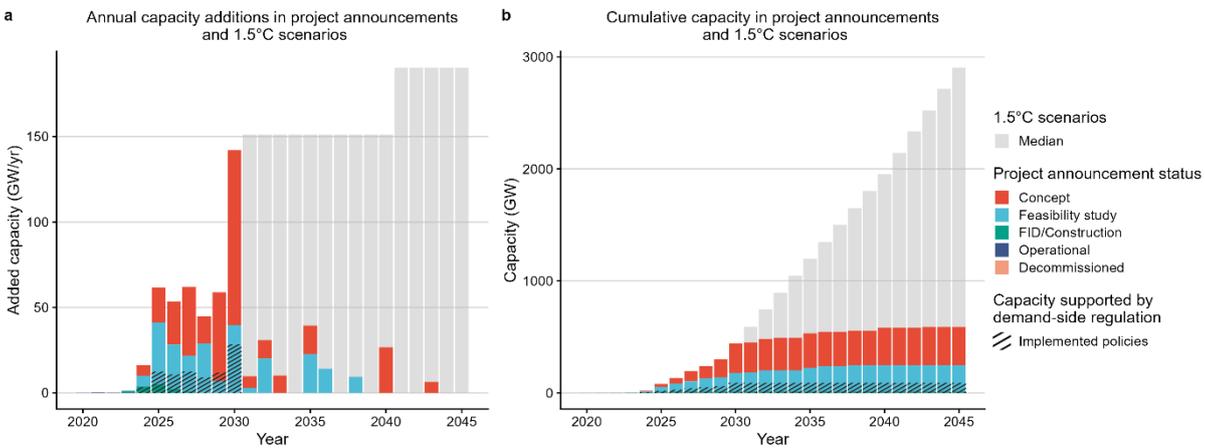

*Extended Data Figure 3: Green hydrogen project announcements and 1.5°C scenario median.* Coloured bars show project announcements by status, while grey bars indicate the median 1.5°C scenario. Hatched bars indicate the part of capacity that is supported by demand-side regulation such as end-use quotas in the central estimate (see Methods). **a,** Annual capacity additions of project announcements and required annual capacity additions for the median of 1.5°C scenarios. For simplicity, we assume a linear scale-up between the project pipeline in 2030, the 1.5°C scenario median in 2040, and the 1.5°C scenario median in 2050 (see Extended Data Figure 2). **b,** Cumulative capacity of project announcements, equivalent to the right bars in Figure 2, and required capacity for the median of 1.5°C scenarios.

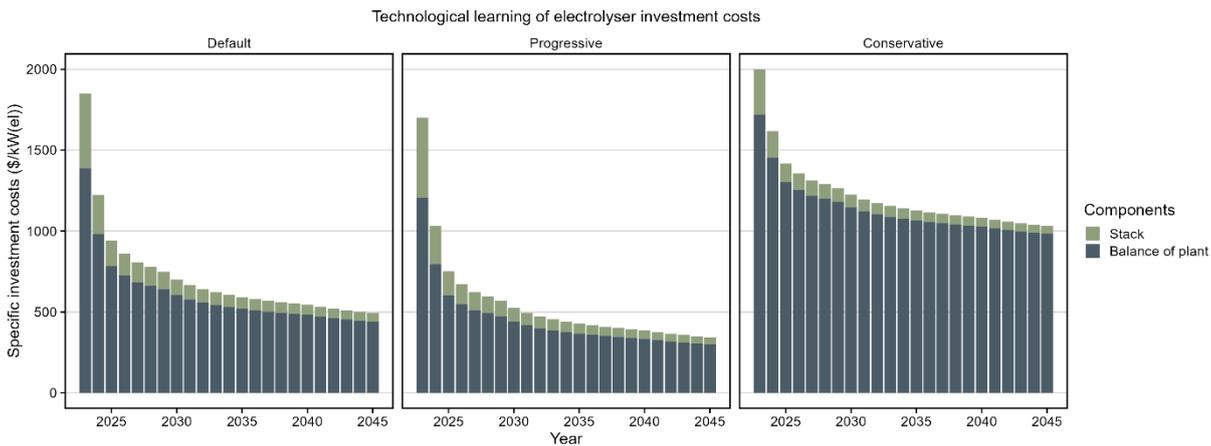

*Extended Data Figure 4: Specific investment costs of electrolysers.* Investment costs decrease due to technological learning (see Methods), with different learning rates for the electrolyser stack and the balance of plant (see Table 1). Until 2030, technological learning is driven by cumulative project announcements, assuming that all projects are realised on time. After 2030, technological learning is driven by the capacity required in the median 1.5°C scenario (see Extended Data Figure 3), assuming a linear scale-up between 2030-2040 and 2040-2050 for simplicity.

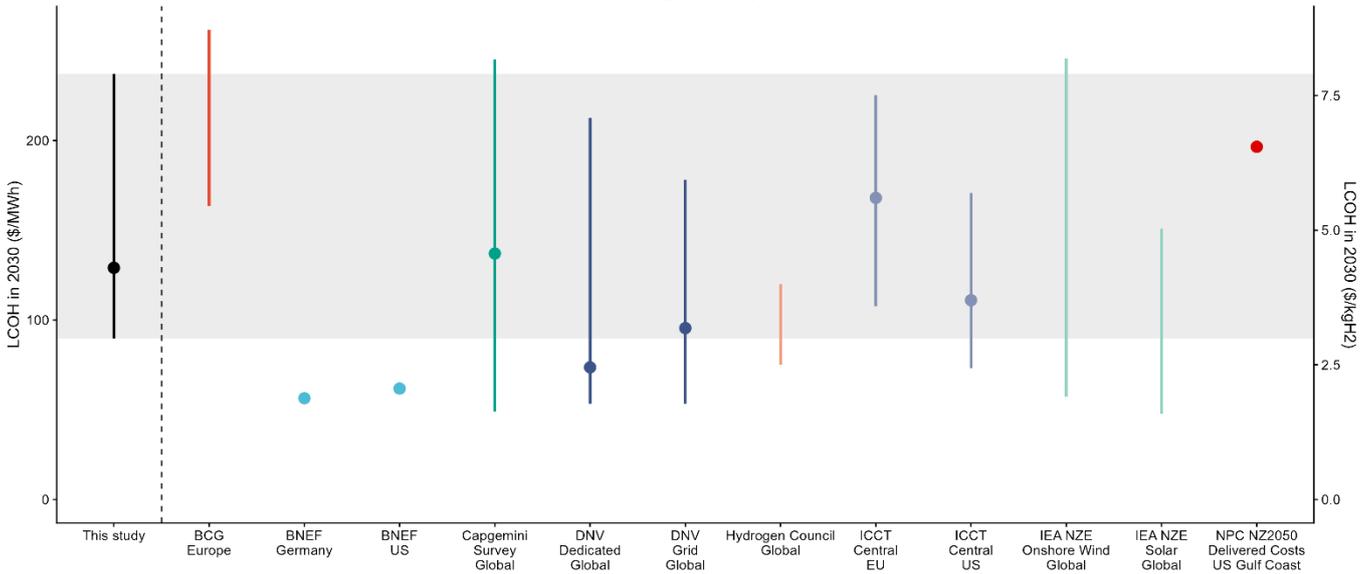

***Extended Data Figure 5: Levelised cost of green hydrogen in 2030 compared to recent studies.*** *The filled circle indicates the central estimate (if available) and the vertical line indicates the range (if available). Colours indicate the organisation. Our 2030 LCOH estimates are in line with most recent studies, but do not extend as far down as some other analyses. We attribute this to (i) recent cost increases of electrolysers, which we include in our calculation, and (ii) a lack of consensus about which parameters should be included in the LCOH calculation[22]. We suspect that studies reporting low LCOH values do not include e.g. transport and storage costs, which are critical for a full assessment of hydrogen costs in our case. Note that Capgemini did not calculate LCOHs, but rather conducted a global survey among more than 100 companies in the hydrogen industry. The data for this figure, including sources and the full name of the organisations, is available on GitHub.*

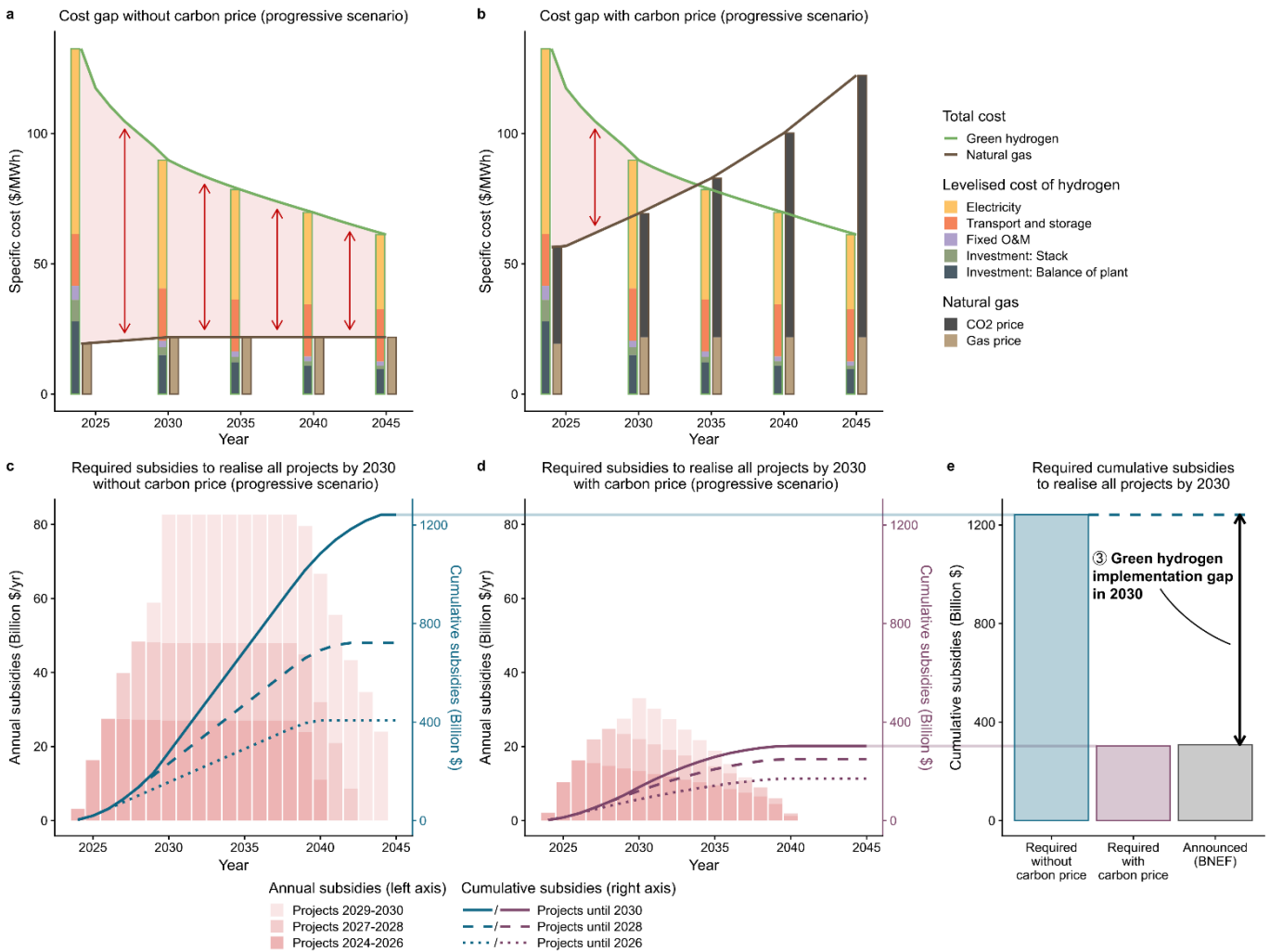

***Extended Data Figure 6: The green hydrogen implementation gap in 2030 (progressive scenario).*** *Analogous to Figure 5 using parameters that are more favourable for green hydrogen (see Table 1 and Methods). **a-b**, Cost gap between the levelised cost of green hydrogen (LCOH) and the price of natural gas without carbon pricing (a) and with carbon pricing (b) for the progressive parameter estimate (see Methods). The red double-headed arrows and the light red shading indicate the cost gap that needs to be bridged by subsidies. The stacked bars indicate the*

*decomposition of the LCOH and the total cost of natural gas for selected years (2024, 2030, 2035, 2040, 2045). For better visibility, the LCOH bar is slightly shifted to the left, and the natural gas bar to the right. **c-d,** Required subsidies to bridge the cost gap without carbon pricing (c) and with carbon pricing (d) in order to realise all project announcements until 2030 on time (see Extended Data Figure 3). The bars show required annual subsidies (left axis), while the lines show required cumulative subsidies (right axis). **e,** Required cumulative subsidies to realise all project announcements until 2030 (without and with carbon pricing) in comparison to globally announced hydrogen subsidies as of September 2023 from BloombergNEF (BNEF)[50]. When calculating subsidies, we take currently implemented demand-side policies, which reduce pressure on supply-side subsidies, into account (see Methods and Extended Data Figure 3). Without carbon pricing, more than $1.2 trillion of subsidies are required to realise all projects announced until 2030. With ambitious carbon pricing, the 2030 green hydrogen implementation gap closes as currently announced subsidies match required subsidies. Note that c-e only show subsidies required for green hydrogen project announcements until 2030. Staying on a 1.5°C scenario requires substantial further subsidies after 2030, which, without carbon pricing, would need to be permanent (see Extended Data Figure 8).*

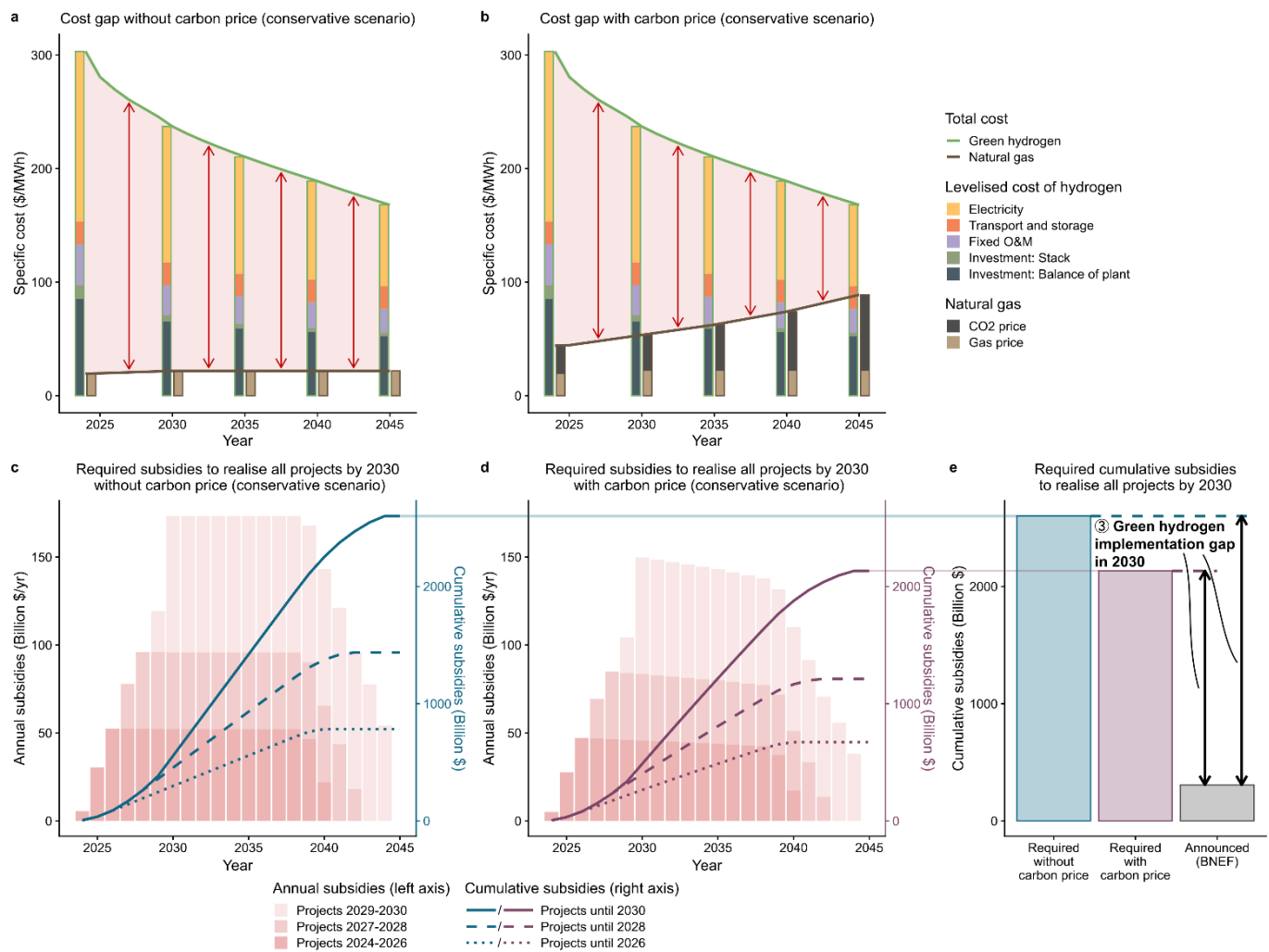

**Extended Data Figure 7: The green hydrogen implementation gap in 2030 (conservative scenario).** *Analogous to Figure 5 using parameters that are less favourable for green hydrogen (see Table 1 and Methods). **a-b,** Cost gap between the levelised cost of green hydrogen (LCOH) and the price of natural gas without carbon pricing (a) and with carbon pricing (b) for the conservative parameter estimate (see Methods). The red double-headed arrows and the light red shading indicate the cost gap that needs to be bridged by subsidies. The stacked bars indicate the decomposition of the LCOH and the total cost of natural gas for selected years (2024, 2030, 2035, 2040, 2045). For better visibility, the LCOH bar is slightly shifted to the left, and the natural gas bar to the right. **c-d,** Required subsidies to bridge the cost gap without carbon pricing (c) and with carbon pricing (d) in order to realise all project announcements until 2030 on time (see Extended Data Figure 3). The bars show required annual subsidies (left axis), while the lines show required cumulative subsidies (right axis). **e,** Required cumulative subsidies to realise all project announcements until 2030 (without and with carbon pricing) in comparison to globally announced hydrogen subsidies as of September 2023 from BloombergNEF (BNEF)[50]. When calculating subsidies, we take currently implemented demand-side policies, which reduce pressure on supply-side subsidies, into account (see Methods and Extended Data Figure 3). Without carbon pricing, more than $2.6 trillion of subsidies are required to realise all projects announced until 2030. With an ambitious carbon price, $2.1 trillion of subsidies would still be required. Note that c-e only show subsidies required for green hydrogen project announcements until 2030. Staying on a 1.5°C scenario requires substantial further subsidies after 2030, which, without carbon pricing, would need to be permanent (see Extended Data Figure 8).*

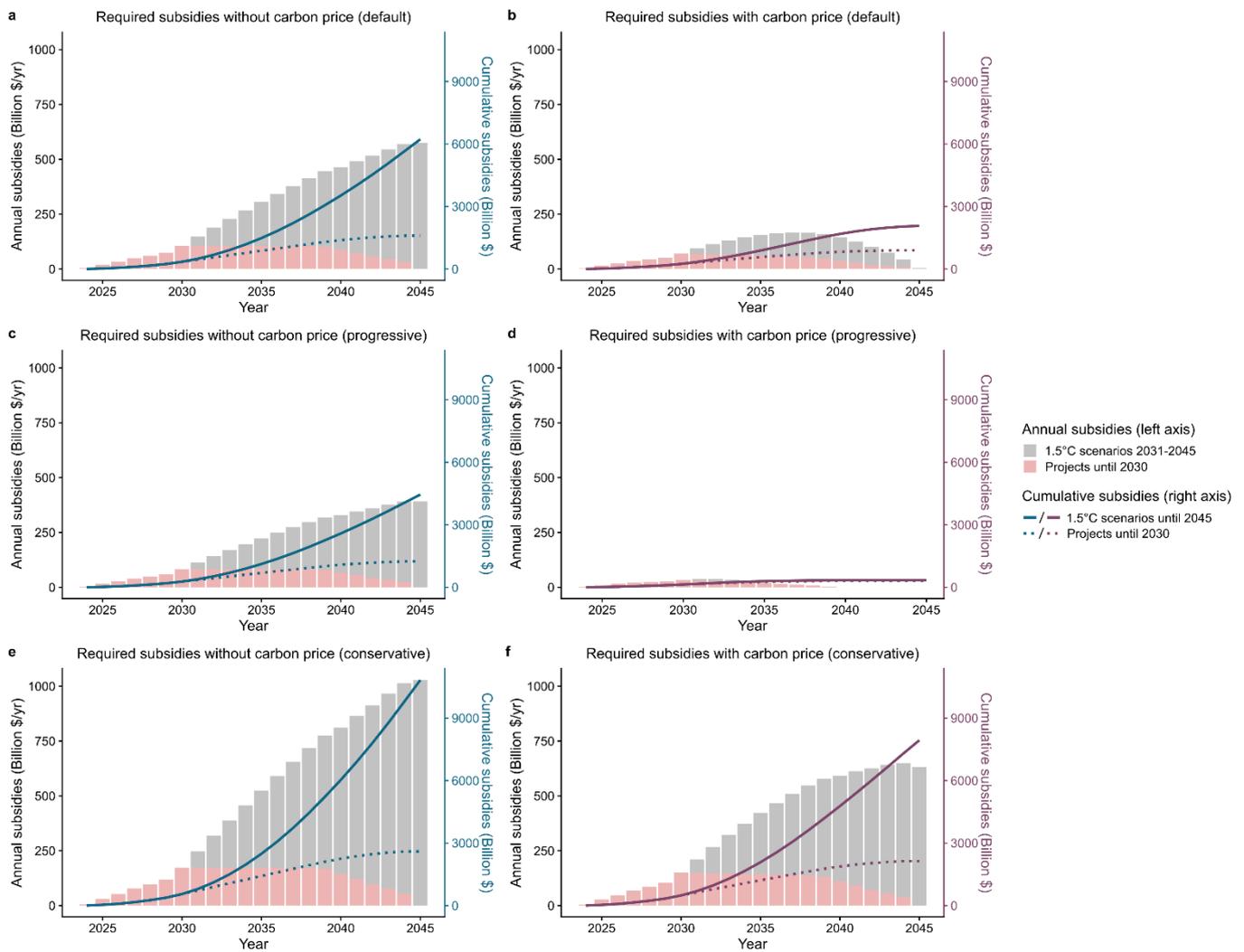

*Extended Data Figure 8: Required subsidies to realise all project announcements until 2030 and follow a median 1.5°C scenario after 2030.*
*Required annual subsidies are displayed as bars, with subsidies for project announcements in light red (compare Figure 5, Extended Data Figure 6 and Extended Data Figure 7) and subsidies for the median 1.5°C scenario in grey (compare Extended Data Figure 2 and Extended Data Figure 3). Required cumulative subsidies are displayed as lines, with dotted lines indicating cumulative subsidies for project announcements until 2030, and solid lines indicating cumulative subsidies for project announcements until 2030 and subsequently for the median 1.5°C scenario.* ***a,c,e,*** *Without carbon pricing, for the central (a), progressive (c), and conservative (e) scenario. Without carbon pricing, the cost gap does not close until 2045, leading to a need for permanent subsidies. In the worst-case scenario, cumulative subsidies required for 1.5°C exceed $10 trillion by 2045 (e).* ***b,d,f,*** *With carbon pricing, for the central (b), progressive (d), and conservative (f) scenario.*

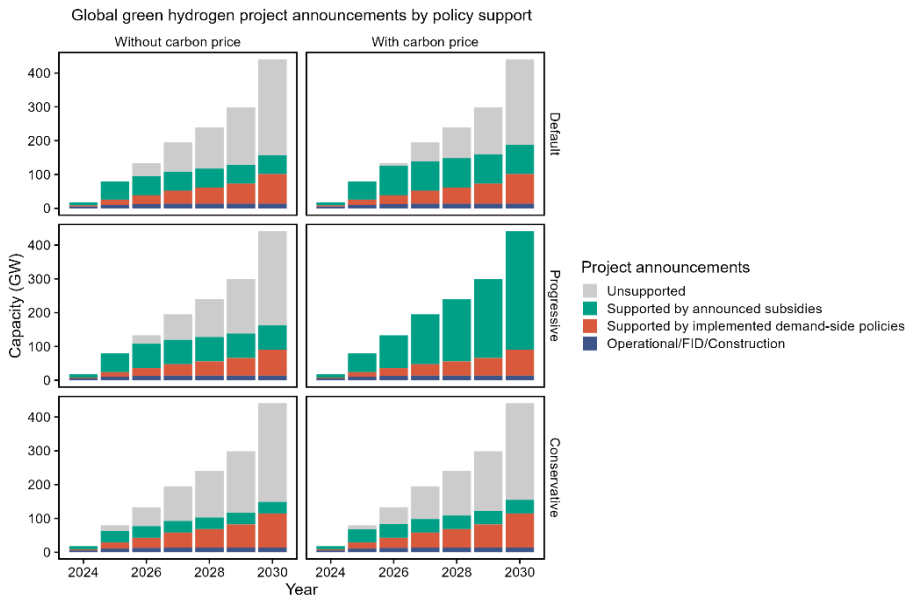

*Extended Data Figure 9: Global green hydrogen project announcements supported by policies.* This figure relies on the same model calculation as in Figure 5 to determine cumulative required subsidies, but addresses the reciprocal question of how much electrolysis deployment would be possible given implemented demand-side policies, estimated at 7 MtH2/yr by 2030[1], and globally announced supply-side subsidies, estimated at $308 billion as of September 2023[50]. On the demand side, we convert the production volume of 7 MtH2/yr to corresponding electrolysis capacity, and distribute it from 2024-2030 in proportion to project announcements (see Methods and Extended Data Figure 3a). Depending on the scenario, implemented demand-side policies may support more capacity than announced supply-side subsidies, underlining the key role of complementing subsidies with demand-side regulation (see Discussion and conclusion).